\documentclass{articlek}

\renewcommand{\refname}{REFERENCES}
\def\be{\begin{equation}}
\def\ee{\end{equation}}

\def\BibTeX{{\rm B\kern-.05em{\sc i\kern-.025em b}\kern-.08em
            T\kern-.1667em\lower.7ex\hbox{E}\kern-.125emX}}

\usepackage{graphicx}

\begin{document}
\sloppy
\twocolumn[{
\vspace*{1.7cm}   
\begin{center}
{\large\bf TERNARY-SPIN ISING MODEL ON AN ANISOTROPICALLY DECORATED SQUARE LATTICE:
           AN EXACTLY SOLVABLE CASE}\\

{\small L. \v{C}anov\'a, lucia.canova@upjs.sk, J. Stre\v{c}ka, jozef.strecka@upjs.sk 
        and J. Dely, jan.dely@upjs.sk, \\
        Department of Theoretical Physics and Astrophysics, Faculty of Science, \\
        P. J. \v{S}af\'arik University, Park Angelinum 9, 040 01 Ko\v{s}ice, Slovak Republic}\\

\end{center}
\vspace*{1ex}

{\bf ABSTRACT.} Magnetic properties of a ternary-spin Ising model on the decorated square lattice are studied within a generalized decoration-iteration transformation. Depending on the mutual ratio between exchange interactions and the single-ion anisotropy, there apprear six different phases in the ground state. The magnetic order of these phases together with the critical behaviour and corresponding magnetization curves are discussed in detail.\\
}]
\section{INTRODUCTION}

Decorated ferrimagnetic Ising models have enjoyed an immense research interest for many years. Besides exactly solvable two-sublattice Ising model \cite{JA_98}, much effort has been recently devoted to the decorated Ising square lattice composed of three magnetic atoms of different magnitudes \cite{KA_02}. However, it is noteworthy that only the compensation phenomenon of this model has been exactly examined yet, while the critical behaviour and thermodynamic quantities have been calculated with Bethe-Peierls approximation \cite{KA_02}. Owing to this fact, the aim of this work is to show a simple way for obtaining complete exact results for this system.

\section{MODEL AND METHOD}

\begin{figure} [t, h]
\begin{center}
\includegraphics[width=69mm]{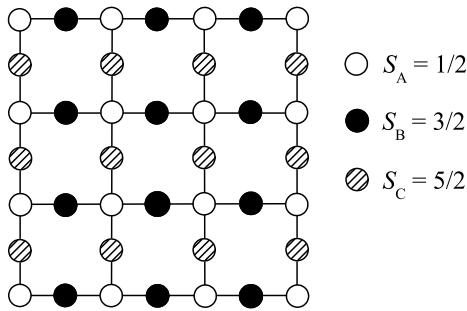} 
\end{center} 
\vspace{-8mm} 
\caption{Part of the decorated Ising square lattice. The white circles denote A atoms, while the black and hatched ones represent decorating B and C atoms, respectively.}
\label{fig1}
\end{figure}

In this article, we shall consider a ternary-spin Ising model on the decorated square lattice (see Fig.~\ref{fig1}), where the lattice sites labeled by white circles are occupied by A atoms with the spin $S_{\rm A} = 1/2$ and the decorating sites denoted by black and hatched circles are occupied by B and C atoms with the spins $S_{\rm B} = 3/2$ and $S_{\rm C} = 5/2$, respectively. The total Hamiltonial of the system reads:
\be
{\cal{H}} 
= \sum_{\langle i,m \rangle} J_{im} S_{i}^{z} S_{m}^{z}
+ \sum_{\langle i,n \rangle} J_{in} S_{i}^{z} S_{n}^{z} 
- D \sum_{k} \left( S_{k}^{z} \right)^2 \,,
\label{eq:H}
\ee
where the first two summations are carried out over the nearest-neighbour A$-$B and A$-$C pairs,
respectively, and the last one runs over the decorating B and C sites. In Eq.~(\ref{eq:H}), $S_{i}^{z}$ and $S_{k}^{z}$ ($k = m, n$) denote the Ising spin variables of A atoms and B or C atoms depending on whether $k$ is selected as $k = m$ or $k = n$. Next, $J_{ik}$ stands for the exchange interaction between the nearest A$-$B neighbours if $k = m$ and the nearest A$-$C neighbours if $k = n$. Finally, the parameter $D$ stands for the single-ion anisotropy acting on B and C atoms. To proceed further, we introduce the generalized decoration-iteration transformation \cite{FI_59}:
\begin{eqnarray}
\sum_{n = -S_k}^{S_k}\hspace{-0.2cm}\exp\left(\beta D n^2\right) 
\cosh\left[\beta n J_{ik}\left(S_{i}^{z}+ S_{i+1}^{z}\right)\right] \nonumber \\
= P_{\alpha}\exp\left(\beta R_\alpha S_{i}^{z}S_{i+1}^{z}\right),\hspace{0.4cm}
\textrm{for}\hspace{0.4cm}\alpha = h, v
\label{eq:DIT}
\end{eqnarray}
where $\beta = 1/(k_{\rm B}T)$ and the subscript $\alpha$ specifies the bond of the lattice to which the transformation (\ref{eq:DIT}) is applied. Namely, Eq.~(\ref{eq:DIT}) is used for horizontal bonds if $k = m$ and then one puts ${\alpha} = h$, while for $k = n$ it is applied for vertical bonds and thus ${\alpha} = v$. To simplify further calculation, we shall assume that interactions $J_{im}$ and $J_{in}$ take exclusively positive values $J_{\rm AB}$ and $J_{\rm AC}$, respectively. By the use of Eq.~(\ref{eq:DIT}), it is then quite straightforward to derive the equality:
\be
{\cal{Z}}(\beta, J_{\rm AB}, J_{\rm AC}, D) = P^{N} {\cal{Z}}_{0}(\beta, R_h, R_v)\,,
\label{eq:Z}
\ee
which relates the partition function ${\cal{Z}}$ of the considered model and that one ${\cal{Z}}_{0}$ of the original anisotropic spin-$1/2$ Ising square lattice with nearest-neighbour couplings $R_{h}$ and $R_v$ in horizontal and vertical direction, respectively. In above, $N$ denotes the total number of A atoms. From the Eq.~(\ref{eq:Z}) one can readily calculate all important quantities utilizing the standard thermodynamic relations. In the present article, we shall turn our attention to the analysis of the ground state and the critical behaviour. In addition, it is also useful to investigate the sublattice magnetization $m_{\rm A} = \langle S_{i}^z \rangle\,, m_{\rm B} = \langle S_{m}^z \rangle\,, m_{\rm C} = \langle S_{n}^z \rangle$, where $\langle \ldots \rangle$ stands for the standard canonical average, and the total magnetization reduced per one atom of the original Ising model $m = M/N = m_{\rm A} + m_{\rm B} + m_{\rm C}$, as well. Finally, let us mention a critical condition, which follows from Onsager's exact solution \cite{ON_44}:
\be
\sinh(\beta_{\rm c}R_h/2)\sinh(\beta_{\rm c}R_v/2) = 1,
\label{eq:Tc}
\ee
where $\beta_{\rm c} = 1/(k_{\rm B}T_{\rm c})$ and $T_{\rm c}$ denotes the critical temperature of the model system.

\section{RESULTS AND DISCUSSION}

\begin{figure} [t, h]
\begin{center}
\includegraphics[width=67.0mm,height=54.0mm]{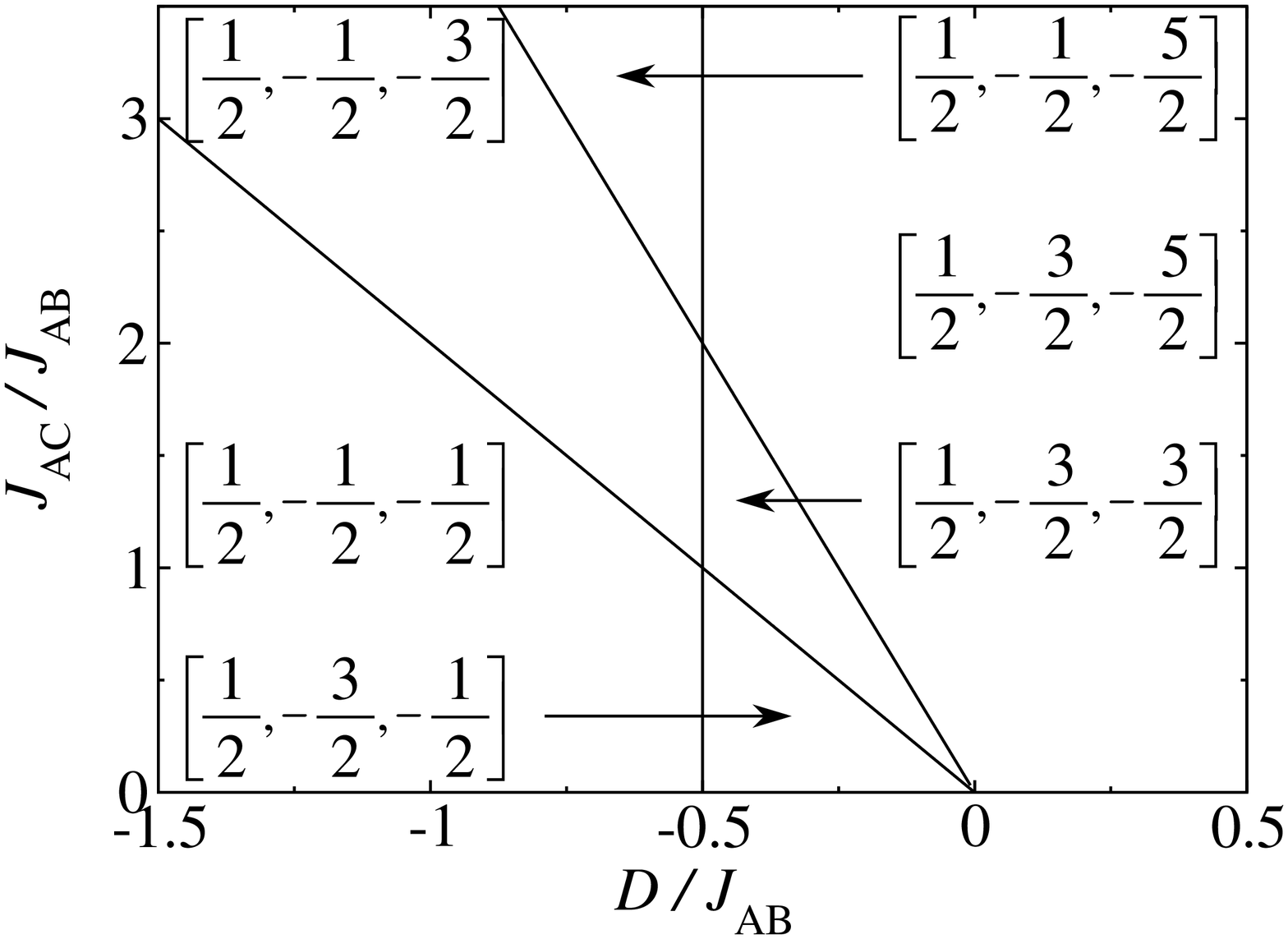} 
\end{center} 
\vspace{-5mm} 
\caption{The ground-state phase diagram.}
\label{fig2}
\end{figure}

Now, we proceed to a brief discussion of the most interesting numerical results obtained for the ground-state and finite-temperature phase diagram. Directly from Fig.~\ref{fig2}, where the spin order drawn in square brackets shows a typical spin configuration $\left[ S_{\rm A}, S_{\rm B}, S_{\rm C} \right]$ in the ground state, one easily finds that A atoms remain in the spin state $S_{\rm A} = 1/2$ in the whole parameter space, while the spin order of decorating B and C atoms is basically modified by the single-ion anisotropy $D$ and the ratio $J_{\rm AC}/J_{\rm AB}$. Actually, a first-order transition $D^{\star} = -0.5 J_{\rm AB}$ associated with the spin crossover of B atoms $S_{\rm B} = -1/2 \leftrightarrow -3/2$ occurs regardless of $J_{\rm AC}/J_{\rm AB}$ as $D$ reinforces. On the other hand, the $J_{\rm AC}/J_{\rm AB}$ strengthening clearly causes spin changes of C atoms from the lowest $S_{\rm C} = -1/2$ towards the highest $S_{\rm C} = -5/2$ available spin state. Therefore, two further first-order transitions $J_{\rm AC}^{\star} = -2 D$ and $J_{\rm AC}^{\star\star} = -4 D$ corresponding to spin crossovers $S_{\rm C} = -1/2 \leftrightarrow -3/2$ and  $S_{\rm C} = -3/2 \leftrightarrow -5/2$, respectively, can be found within the $D < 0.0$ region. In consequence of this, there appear in total six different ferrimagnetically ordered phases in the ground state depending on the ratio between $D$,  $J_{\rm AB}$ and $J_{\rm AC}$. 

\begin{figure} [t, h]
\begin{center}
\includegraphics[width=80mm]{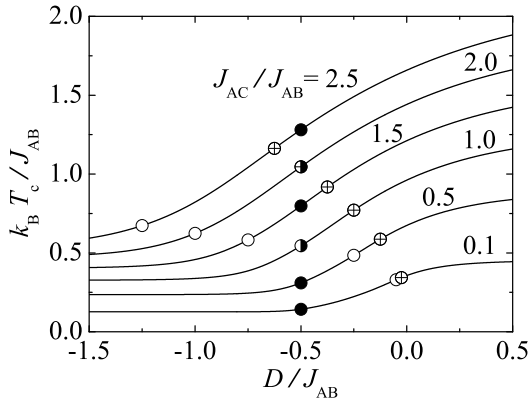} 
\end{center} 
\vspace{-9mm} 
\caption{The critical temperature versus the sigle-ion anisotropy $D$ for several values 
         of the ratio $J_{\rm AC}/J_{\rm AB}$.}
\label{fig3}
\end{figure}

Next, let us make some comments on the finite-temperature phase diagram displayed in Fig.~\ref{fig3}, which shows the variation of critical temperature with the single-ion anisotropy $D$ for several values of the\hspace{-0.3cm}
\vspace{-5mm} 
\begin{figure} [t, h]
\begin{center}
\includegraphics[width=64.0mm,height=54.0mm]{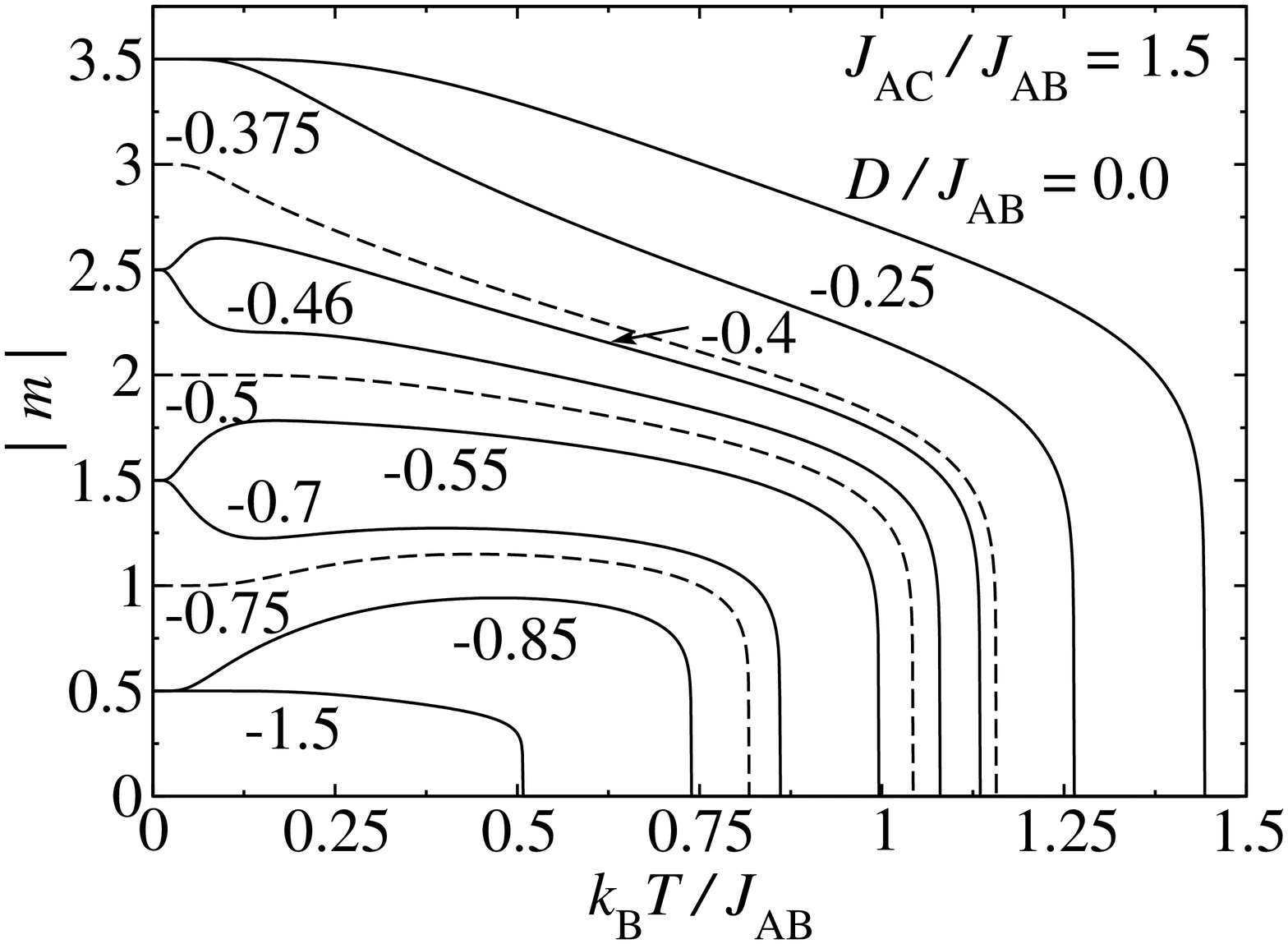} 
\end{center} 
\vspace{-5mm} 
\caption{The thermal variations of the total magnetization for $J_{\rm AC}/J_{\rm AB} = 1.5$ 
         and several values of $D$.}
\label{fig4}
\end{figure}
ratio $J_{\rm AC}/J_{\rm AB}$. As one can see, the critical temperature monotonically decreases from its maximum value at $D \rightarrow \infty$, and tends towards its global minimum achieved in the limit $D \rightarrow -\infty$ regardless of the ratio $J_{\rm AC}/J_{\rm AB}$. Thus, the spin system remains long-range ordered in the ground state independently of the $J_{\rm AC}/J_{\rm AB}$ strength in agreement with the previous ground-state analysis. Open, crossed and black circles denote in Fig.~\ref{fig3} those critical temperatures at which spin crossovers $S_{\rm C} = -1/2 \leftrightarrow -3/2$, $S_{\rm C} = -3/2 \leftrightarrow -5/2$ and $S_{\rm B} = -1/2 \leftrightarrow -3/2$, respectively, emerge in the ground state. Besides, symbols arising from mixing of relevant circles represent special critical points that correspond to the coexistence of four different ferrimagnetic phases (compare Figs.~\ref{fig2} and \ref{fig3}).

In Fig.~\ref{fig4}, we show some typical thermal variations of the total magnetization for $J_{\rm AC}/J_{\rm AB} = 1.5$ and several values of $D$. As one can see, the plotted results are completely consistent with those presented in Figs.~\ref{fig2} and \ref{fig3}. Indeed, the initial value of $|m|$ takes one of four possible values $|m| = 1/2, 3/2, 5/2$ or $7/2$ corresponding to $[\frac{1}{2}, -\frac{1}{2}, -\frac{1}{2}]$, $[\frac{1}{2}, -\frac{1}{2}, -\frac{3}{2}]$, $[\frac{1}{2}, -\frac{3}{2}, -\frac{3}{2}]$ or $[\frac{1}{2}, -\frac{3}{2}, -\frac{5}{2}]$ phases, respectively. Furthermore, there are also three special cases starting at $|m| = 1, 2$ and $3$ (labeled by dashed lines in Fig.~\ref{fig4}), which correspond to two-phase coexistence at $T = 0$.

\section{CONCLUSIONS}

In conclusion, it should be mentioned that the model under investigation shows rich critical behaviour and many other interesting features, which require further detailed analysis. Another interesting results concerning this model will be therefore presented in the more comprehensive review to be published in the near future.
\\

\noindent ACKNOWLEDGMENT: This work was financially supported by the grant VVGS 12/2006.


\begin{thebibliography}{99}

\leftskip=-5pt \vspace{-0.3truecm}
\bibitem{JA_98} M.~Ja\v{s}\v{c}ur, Physica A {\bf 252}, 217 (1998).
\bibitem{KA_02} T.~Kaneyoshi, Physica A {\bf 303}, 507 (2002).
\bibitem{FI_59} M.~E.~Fisher, Phys. Rev. {\bf 113}, 969 (1959).
\bibitem{ON_44} L.~Onsager, Phys. Rev. {\bf 65}, 117 (1944).

\end{thebibliography}
\end{document}